  \documentstyle [twoside,12pt,epsfig]{article}  
  \def\ZzZ{{\hbox{\tenrm Z\kern-.31em{Z}}}} 
  \def\CcC{{\hbox{\tenrm C\kern-.45em{\vrule height.67em width0.08em depth- 
  .04em 
  \hskip.45em }}}} 
   
  \def\mapbelow#1{\smash{\mathop{\longrightarrow}\limits_{#1}}} 

  \newcommand{\lab}{\label}

  \newcommand{\bc}{\begin{center}} 
  \newcommand{\ec}{\end{center}} 
  \newcommand{\be}{\begin{equation}} 
  \newcommand{\ee}{\end{equation}} 
  \newcommand{\bea}{\begin{eqnarray}} 
  \newcommand{\eea}{\end{eqnarray}} 
  \newcommand{\bs}{\begin{subequations}} 
  \newcommand{\es}{\end{subequations}} 
  \newcommand{\beq}{\begin{eqalignno}} 
  \newcommand{\eeq}{\end{eqalignno}} 

  \newcommand{\half}{\frac{1}{2}} 
  \newcommand{\qrt}{\frac{1}{4}} 
   
  \def\PRD{{ Phys. Rev.}   D} 
   
  \def\om{\omega} 
  \def\Om{\Omega} 
   
  \def\lab{\label} 
   
\begin{document} 
   
  \title{Canonical quantization and  expanding metrics}
   
  \author{Eleonora Alfinito\thanks{E-mail:alfinito@pcvico.csied.unisa.it} ~and 
  Giuseppe Vitiello\thanks{E-mail:vitiello@pcvico.csied.unisa.it}
   \\
  \and
  { \it Dipartimento di Fisica, Universit\`a di Salerno, 84100 } \\
  {\it and INFN Gruppo Collegato di Salerno }  }
  
  \date{}
  
  \maketitle

  \begin{abstract} The canonical formalism 
  for expanding metrics scenarios is presented.
  Non-unitary time evolution implied by expanding 
  geometry is described as a trajectory over  
  unitarily inequivalent
  representations at different times of the canonical commutation 
  relations.
  Thermal properties of inflating Universe are also discussed. 
  \end{abstract}
  
  \normalsize

  \bigskip
   
  
  \bigskip
   
  In this paper we study the quantum field theory (QFT) formalism for  
  non-unitary time evolution implied by  
  inflationary models and  
  in general by expanding metrics scenarios. As well known, non-unitary  
  time evolution cannot be handled in the canonical formalism  
  and for this reason many semi-classical techniques are currently  
  used; the problem of the canonical  
  quantization of inflationary time evolution is thus still at  
  an unsatisfactory stage.  
  In this paper we show that the operator formalism for  
  the canonical quantization of time 
  evolution in expanding geometry can be formulated provided the  
  full set of unitarily inequivalent  
  representations of the canonical commutation relations are  
  considered. 
   
  Let us start our discussion, which will be limited to gravitational  
  wave modes, by denoting, as customary, the 
  flat 
  time-dependent metrics by $g_{\mu\nu}(t)=g^{0}_{\mu\nu}(t)  
  + h_{\mu\nu}(t)$. As well known, use of the De Donder gauge condition  
  $\partial_{\mu} h_{\mu\nu}\,=\,0$ and the Einstein  
  equations give 
  \be\Box \ h_{\mu\nu}\,=\,0.\ee 
  The field $h_{\mu\nu}$ may be then decomposed into harmonic modes  
  $u_{k}$ 
  obeying the equation: 
  \be 
  \ddot{u}_{ k}(t)+H \dot{u}_{ k}(t) 
  +\omega_{ k} ^2(t)u_{ k}(t)=0   
  \lab{p12}\ee 
  with   
  \be 
  {{\omega_{k}}^2}(t)=\frac{k^{2}c^2}{a^2(t)},\qquad a(t)\,=\,a_{0} 
  e^{\frac{1}{3}Ht} \quad .
  \lab{p14}\ee 
  $H$ is the Hubble constant.
  In the Minkowski 
  space-time $\om_{k}$ is constant in time, but when the Universe expands,  
  $\om_{k}$ is time-dependent: $\om_{k} = \om_{k}(t)$.  
   
  In the following, where not strictly necessary, we will omit 
  the  $ k$-index, remembering that each equation is written  
  down for fixed $ k$. 
   
  The first order derivative term $H \dot{u_{ k}}$ in eq.(\ref{p12}) is  
  generally incorporated into the frequency term by using the  
  conformal time variable $\eta$ 
  \cite{Grish1, Grish2, Grish3}; such a 
  computational strategy is very  
  useful in the phenomenological approach, however our purpose in  
  this paper is to illustrate the subtleties of the canonical  
  quantization for non-unitary  
  time evolution and therefore we must explicitly take care of 
  the  
  inflation term in eq.(\ref{p12}). In this way the full structure of  
  the state space will be revealed. 
   
  Of course, it is the term $H {\stackrel{.}u}_{k}$ in (\ref{p12})  
  which makes impossible to proceed to canonical quantization (as a matter  
  of fact, it is not even possible to define a canonical conjugate  
  momentum to the $u$ variable in (\ref{p12})).  
  However, as we will show in the following, the {\it double}  
  oscillator system  
  \be\stackrel{..}{u}+H \stackrel{.}{u}+\omega ^2(t)u=  0 ~,\lab{p20}\ee  
  \be \stackrel{..}{v}-H \stackrel{.}{v}+\omega ^2(t)v=  0 ~, 
  \lab{p21} 
  \ee 
  does admit a canonical quantization procedure, provided one works in the
  QFT framework\cite{FT,Bat,QD}.  
  Note that 
  in the same way as the $u$ oscillator describes the 
  expanding (inflating)  
  metrics, the oscillator $v$ can be associated to the  
  "contracting" ("deflating") metrics (in this sense we might  
  speak of a "double Universe"\cite{Double,Jagna}):  
  The $u-v$ system is a non-inflating (and non-deflating) system.  
  This is why it is now possible to set up  
  the canonical quantization scheme. 

  In order to better understand the need to double the degrees of freedom,  
  it is useful to observe that $v = u e^{Ht}$ is solution of (\ref{p21}) 
  and that by setting $u(t) = {1\over {\sqrt 2}} r(t)e^{{-Ht\over 2}}$ and  
  $v(t) = {1\over {\sqrt 2}} r(t)e^{{Ht\over2}}$ the system of equations 
  (\ref{p20}) and (\ref{p21}) is equivalent to the single  
  parametric oscillator $r(t)$ (see also  
  \cite{BGPV}): $ 
  \stackrel{..}{r} +\Omega ^2(t)r=  0 ~$. 
  
  The physical reason to double the degrees of  
  freedom relies thus in the fact that one must  
  work with closed systems as required, indeed,  
  by the canonical quantization  
  formalism. 
  
  We stress that the doubling of the degrees of  
  freedom is intrinsic to the Bogolubov transformations (see below),  
  so that one deals  
  with a doubled system anytime one works with such transformations.  
  For this  
  reason all the "mixed modes" formalisms (since Parker's  work \cite{Park} ) necessarily involve the algebraic structure of the  
  doubling of the modes.

  It turns out to be convenient to introduce  
  the canonical transformations 
  \be u(t)\ =\  \frac{U(t) + V(t)}{\sqrt 2}, \quad 
  v(t)\ = \ \frac{U(t) - V(t)}{\sqrt 2}. 
   \lab{p35}\ee 
   
  We are thus dealing  
  with the decomposition of the parametric oscillator $r(t)$ on the {\it  
  hyperbolic} plane (i.e. in the pseudo-Euclidean metrics): $r^{2}(t) =  
  U^{2}(t) -V^{2}(t)$. The Hamiltonian for our coupled oscillator  
  system is \cite{QD, Double}: 
  \be {\cal H}   = {1 \over 2} {p_U}^2 + {1 \over 2}{\Omega}^2(t)U^2 
  -{1 \over 2} {p_V}^2  
    - {1 \over 2}{\Omega}^2(t)V^2   - 
  {\Gamma} ({p_U} V + {p_V} U).\lab{p42}\ee 
  with $\Gamma \equiv {H \over 2}$ and 
  $\Omega(t) \equiv$ $\left [ \left ({\omega}^{2}(t)  - {{H  
  ^{2}}\over{4}} \right ) \right ]^{1\over{2}}$, which we assume to  
  be real (for any $k$ and any $t$) 
  in order to avoid 
  over-damped regime; i.e. we assume 
  ${\Omega}^{2}(t) \ge 0$.  As we will show later on,  
  this condition turns out to act as  
  a cut-off on $k$. 
  
  Now it is possible to introduce  
  the annihilation  operators:  
  \be 
  A  = \frac{1}{\sqrt{2}} \left(\frac{p_U}{\sqrt{\hbar\om_{0}}}-iU 
  \sqrt{\frac{\om_{0}}{\hbar}} 
  \right), \qquad B 
   = \frac{1}{\sqrt{2}} \left(\frac{p_V}{\sqrt{\hbar\om_{0}}}- 
   iV\sqrt{\frac{\om_{0}}{\hbar}} 
  \right),  
  \lab{p44}\ee 
  and the corresponding creation operators with usual commutation relations.
  
  Then it can be shown \cite{QD} that the vacuum state $| 0>$ is unstable:
  \be 
  <0(t) | 0> \, \propto \exp{\left ( - t  \Gamma \right )} 
  \rightarrow 0 \; ~for~large~t , 
  \lab{p712}\ee  
  i.e. time evolution brings "out" of the initial-time Hilbert  
  space for large $t$.  
  This is not acceptable in quantum mechanics since there the Von Neumann  
  theorem states that all the representations of the canonical commutation  
  relations are unitarily equivalent and therefore, as already remarked 
  above, there is no room in  
  quantum mechanics for 
  non-unitary time evolution as the one in (\ref{p712}). 
  On the contrary, in QFT there exist infinitely  
  many unitarily inequivalent  
  representations and this leads to us to study 
  our problem in the framework of QFT. 
   
  To set up  
  the formalism in QFT we have to consider the infinite volume limit;  
  however, as customary, we will work at finite volume and at the end of  
  the computations we take the limit $V \rightarrow \infty$. The QFT  
  Hamiltonian  is 
  \be 
  {\cal H}  = {\cal H}_{0} + {\cal H}_{I_1} + {\cal H}_{I_2} 
  \lab{ph1}\ee 
  \be 
  {\cal H}_{0} =  \sum_{ k} \half \hbar \Om_{0,{ k}}(t) 
  (A^{\dagger}_{ k} A_{ k} - B^{\dagger}_{ k} B_{ k} ) 
  \equiv  \sum_{ k} \hbar \Om_{0,{ k}}(t){\cal C}_{ k} ~, 
  \lab{ph2}\ee 
  \be 
  {\cal  H}_{I_1} = - \sum _{ k}{1\over 4}\hbar \Om_{1,{ k}}(t) 
  \left[\left(A_{ k}^{2} 
  + {A^{\dagger}_{ k}}^{2} \right) - \left( B_{ k}^{2} + {B^{\dagger}_{  
  k}}^{2} \right)\right] 
  \equiv  - \sum_{ k} \hbar \Om_{1,{ k}}(t)K_{1,{ k}} ~, 
  \lab{ph3}\ee 
  \be 
  {\cal H}_{I_2} = i\sum _{ k} {\Gamma}_{ k} 
  {\hbar} \left(A^{\dagger}_{ k} B^{\dagger}_{ k} -A_{ k}B_{ k}  
  \right) \equiv i\sum_{ k} \hbar {\Gamma_{ k}} 
  ( J_{+,{ k}} - J_{-,{ k}}) ~, 
  \lab{ph4}\ee 
  with  $\Omega_{0,1,k}= \om_{0}  
  \left(\frac{{{\Om}_k}^{2}(t)}{\om^{2}_{0}}\pm 1\right) ~~$ and 
  \be 
  [A_{ k},A^{\dagger}_{{ k}'}]  =  {\delta}_{{ k},{ k}'} =  
   [B_{ k},B^{\dagger}_{{ k}'}],\qquad    
  {[}A_{ k},B_{{ k}'}{]}  =  0\,=\,  
   [A^{\dagger}_{ k},B^{\dagger}_{{ k}'}] . 
  \lab{p51}\ee 
  
  The group structure underlying the 
  Hamiltonian (\ref{ph1}) is the one of SU(1,1); in fact, the  
  operators $K_{0,k} \equiv {{\cal C}_{k}}$, $K_{1,k}$ and $K_{2,k}\,=\,  
  i\qrt \left[ \left({A_{k}} ^{2} 
   - {{A_{k}}^{\dagger} }^{2}  \right) + \left( {B_{k}} ^{2} 
   -{{B_{k}}^{\dagger} }^{2}  \right)\right] $ close the su(1,1) algebra: 
  \be
  [\, K_{1,k} , K_{2,k}\, ]  = -i K_{0,k},\quad [\, K_{2,k}  , K_{0,k}\, ] = i  
  K_{1,k}, \quad [\, K_{0,k}  , K_{1,k}\, ]  =i K_{2,k}. 
  \lab{}\ee 
  Similarly, the operators  $ 
  J_{+,k }= {A_{k}}^{\dagger} {B_{k}}^{\dagger} , \; J_{-,k }=  
  {A_{k}}{B_{k}} , \; 
  J_{0,k }= \frac{1}{2}({A_{k}}^{\dagger} {A_{k}} + 
  {B_{k}}^{\dagger} {B_{k}}  + 1)$ close the su(1,1) algebra  
  \be 
  [\, J_{+,k} , J_{-,k}\, ] = - 2 J_{3,k} \quad , 
  \quad [\, J_{3,k}  , J_{\pm ,k}\, ] = \pm  
  J_{\pm ,k}. \quad  \lab{p72} 
  \ee 
   
  \noindent ${{\cal C}_{k}}$ and $J_{2,k}$ are proportional 
  to the Casimir operators for the algebra 
  generated by the $J$'s operators and the $K$'s operators,  
  respectively. 

  We remark that 
  \be 
  [{\cal H}_{0}, {\cal H}_{I_2}] = 0 = [{\cal H}_{I_1} , {\cal H}_{I_2}]~, 
  \lab{p62}\ee 
  which guarantees that the minus sign appearing in  
  ${\cal H}_{0}$ is not 
  harmful, i.e., once one starts with a positive definite Hamiltonian it  
  remains 
  lower bounded under time evolution.

  By using the transformation ${\cal H} 
  \rightarrow {{\cal H}^{\prime}} \equiv S(\theta)^{-1} {\cal H}
   S(\theta)$ with $ S(\theta) \equiv \prod_{ k} 
  e^{-i\theta_{k}(t)K_{2,k}}$ and 

  \be 
  \tanh {\theta_{k}(t)} =  
  -{\Om_{1,k} (t) \over \Om_{0,k}  
  (t)} 
  \lab{c0}\ee 
  at any $t$ for any given $k$ we can  
   "rotate away" \cite{so} ${\cal H}_{I_1}$: 
  \be 
  {{\cal H}^{\prime}} \equiv S(\theta)^{-1} {\cal H} S(\theta)
   = {\cal H}^{\prime}_{0} + {\cal H}_{I_{2}}~~. 
  \lab{p74}\ee 
  
  Notice that
  \be 
  {{\cal H}^{\prime}}_{0} = \sum_{ k}\hbar \Om_{ k}(t) 
  (A_{ k}^{\dagger} A_{ k} - B_{ k}^{\dagger} B_{ k} )~~,
  \lab{ph5}\ee 
  and $ [{{\cal H}^{\prime}}_{0} , {\cal H}_{I_{2}} ] = 0 $.
  Also, 
  in a consistent way, the modulus of ${\Om_1 (t) \over \Om_0  
  (t)}$  is  less or equal to $1$ for any $t$ for  any  given  $k$.  
  Later we will comment more on this point.  
   
  When the initial state, say at arbitrary  
  initial time $t_{0}$, ($t_{0} =0$, ${\theta}_{k} ({0})\equiv {\theta}_{k}$  
  for sake of simplicity)  
  is the {\it vacuum} $|0>$ for ${{\cal H}^{\prime}}_{0}$, with 
  $A_{k} |0> = 0 =  
  B_{k}|0>$, 
  the state 
  $
  |0(\theta)> = S(\theta) |0> ~ 
  $  is the zero energy eigenstate 
  (the {\it vacuum}) of ${\cal H}_{0} + {\cal H}_{I_1}$ at $t_{0}$: 
  \be 
  ({\cal H}_{0} + {\cal H}_{I_1})|_{t_{0}}|0(\theta )> = S(\theta)
  {\cal H}^{\prime}_{0}|0> = 0. ~~  
  \lab{p77}\ee 
   
  We observe that the 
  operators $A_{k}$ and $B_{k}$ transform under 
  $\exp{\left ( -i{\theta}_{k} K_{2,k} \right )}$  as 
  \be 
  A_{k}  \mapsto A_{k} (\theta) =  {\it e}^{ 
  -i{\theta}_{k} K_{2,k} }A_{k}  {\it e}^{i{\theta}_{k} K_{2,k}} = 
  A_{k}  \cosh{(\half{\theta}_{k} )} + {A_{k}} ^{\dagger} \sinh{( 
  \half{\theta}_{k} )}~, 
  \lab{p7261}\ee
  \be
  B_{k}  \mapsto B_{k} (\theta)  =  {\it e}^{ 
  -i{\theta}_{k} K_{2,k} }B_{k}  {\it e}^{i{\theta}_{k} K_{2,k} } = 
   B_{k}  \cosh{(\half{\theta}_{k} )} + {B_{k}} ^{\dagger} \sinh{( 
  \half{\theta}_{k} )}~. 
  \lab{p7262}\ee
  These transformations 
  are nothing else than the {\em squeezing}  
  transformations and preserve the commutation relations (\ref{p51}). 
  One has  
  $ 
  A_{k} (\theta) |0(\theta)> = 0 = B_{k} (\theta) |0(\theta)> $~.
   
  The state $|0(\theta)>$ is thus the squeezed 
  vacuum (at this level actually it is not, strictly speaking,  
  a {\em squeezed  
  state} since squeezed states are obtained by applying the squeezing  
  generator to a (Glauber-type) coherent state). Thus we recover  
  the squeezing phenomenon in inflating model discussed elsewhere  
  \cite{Grish1, Grish2, Grish3, Al}.  
   
  Using the commutativity of $J_2$ with $K_{2}$, the  
  $t$-evolution of the squeezed vacuum $|0(\theta)>$ is obtained as (at
  finite volume $V$):
  \be 
  |0(\theta,t)> = \prod_{ k} {1\over{\cosh{(\Gamma_{ k} t)}}} \exp{ 
  \left ( \tanh {(\Gamma_{ k} t)} J_{{ k}, +}(\theta) \right )}  
  |0(\theta)> \quad,  \lab{pq79}\ee 
  with $J_{k,+}(\theta) = A^{\dagger}_{k}(\theta)  B^{\dagger}_{k}(\theta)$. 
  We have  $A_{ k}(\theta,t) |0(\theta,t)> = 
  0 = B_{ k}(\theta,t) 
  |0(\theta,t)>$ with 
  \be
  A_{k}(\theta)\mapsto A_{k}(\theta,t)  = {\it e}^{- i {t\over{\hbar}} {\cal  
  H}_{I_2}} A_{k}(\theta) {\it e}^{i {t\over{\hbar}} {\cal H}_{I_2}}  =   
  A_{k}(\theta) \cosh{({\Gamma}_{k} t)} - B_{k}(\theta)^{\dagger} \sinh{( 
  {\Gamma}_{k} t)} \quad , 
  \lab{795}\ee 
  \be 
  B_{k}(\theta) \mapsto B_{k}(\theta,t)  = {\it e}^{- i {t\over{\hbar}} {\cal  
  H}_{I_2}} B_{k}(\theta) {\it e}^{i {t\over{\hbar}} {\cal H}_{I_2}}  =  -  
  A_{k}(\theta)^{\dagger} \sinh{({\Gamma}_{k} t)} + B_{ k}(\theta) \cosh{( 
  {\Gamma}_{k} t)} \quad. \lab{p796}\ee 
  Notice that these are the time-dependent, canonical Bogolubov 
  transformations.
  
  The state $|0({\theta},t)>$ is a normalized state,~ $ 
  <0(\theta,t) | 0(\theta,t)> = 1 \quad \forall t$, ~and  is  a  
  $su(1,1)$  generalized coherent state. 
  Provided ${\sum_{ k}  
  \Gamma_{ k} > 0}$, non-unitary time evolution is now expressed by 
  (cf. (\ref{p712})): 
  \be 
  <0(\theta,t) | 0(\theta)> \, \propto \exp{\left ( - t  \sum_{ k}   
  \Gamma_{ k} \right )} \rightarrow  0 \; ~ for~large~t~ . \lab{pq712}\ee 
  Use of the customary 
  continuous limit relation $ \sum_{ k}  
  \mapsto {V\over{(2 \pi)^{3}}} \int \! d^{3}{{ k}}$,  
  for $\int \! 
  d^{3}{ k} \, \ln \cosh {(\Gamma_{ k} t)}$  
  finite and positive, gives in the 
  infinite volume limit 
  \be 
  <0(\theta,t) | 0(\theta)> \mapbelow{V \rightarrow \infty} 0 \quad \forall  
  \, t \quad , 
  \lab{p713}\ee 
  \be 
  <0(\theta,t) | 0(\theta',t') > \mapbelow{V \rightarrow \infty} 0 \quad  
  with ~~ \theta' \equiv \theta (t_{0}'),~~\forall \, t\, , t'\, , t_{0}'  
  \quad , \quad t \neq t'~~ .  \lab{p714}\ee

  Eqs. (\ref{p713}) and (\ref{p714}) show that in the infinite volume
  limit the vacua at $t$ and at $t'$, for any $t$ and $t'$, are
  orthogonal states and thus the corresponding Hilbert spaces are unitarily
  inequivalent spaces. This means that the set of states of the system
  splits into unitarily inequivalent representations $\{|0(\theta,t)>\}$
  labeled by $t$.
 
  Thus, the result we have obtained is that the system in its 
  evolution runs over a variety of  
  representations labeled by $t$ of the canonical commutation relations
  which are unitarily  
  inequivalent to each other for $t \neq t'$ in the infinite-volume limit:  
  {\it the non-unitary character of time evolution implied by expanding 
  geometry is thus  
  recovered,  in a consistent scheme, in the unitary inequivalence among  
  representations at different times in the infinite volume limit}. 
  
  The number of modes of type $A_{k}(\theta)$ in the state  
  $|0(\theta,t)>$ is given, at each instant $t$ by 
  \be 
  {n}_{A_{k}}(t) \equiv < 0(\theta,t) | A_{k}^{\dagger} 
  (\theta)  
  A_{ k}(\theta) | 0(\theta,t) > = 
   \sinh^{2}\bigl( \Gamma_{ k} t \bigr) \quad , 
  \lab{p274}\ee 
  and similarly for the modes of type $B_{k}(\theta)$.  
   
  We also observe that the commutativity of ${\cal C}$ (i.e. $K_{0}$)  
  with ${\cal  
  H}_{I_2}$ (i.e. $J_{2}$) ensures that the  
  number $\left( n_{A_{k}} - n_{B_{k}}  
  \right)\,$ is a constant of motion for any $ k$ and any $\theta$. 
  Moreover, one can show \cite{QD,TFD} that 
  the {\sl creation} of a mode $A_{k}(\theta)$ is equivalent to the  
  {\sl destruction} of a mode $B_{k}(\theta)$ and vice-versa.  This means  
  that the $B_{k}(\theta)$ modes can be interpreted  as the {\sl  
  holes} for the modes $A_{k}(\theta)$: the $B$-system can be  
  considered as the sink where the energy dissipated  
  by the $A$-system flows. 
   
  Notice that in the continuum limit,  
  as well known, the $A_{ k}$ (and $B_{ k}$) operators are  
  not well defined on vectors in the Fock space; for instance, 
  since $[A_{k},A^{\dagger}_{{ k}'}] = {\delta}({ k} - { k}')$, ~$|A_{ k}>  
  \equiv A_{ k}^{\dagger}|0>$ is not a normalizable vector:  
  $<A_{ k}|A_{ k}> = \delta ({ 0})$  ~
  which is infinity.  
  As customary one must then introduce wave-packet (smeared out) operators  
  $A_{f} = {1 \over{(2\pi)^{3/2}}}\int {d^3{ k}} A_{ k} f({ k}) $, 
  with spatial distribution described by square-integrable  
  (orthonormal) functions $f(x)$. The commutators are 
  \be 
  [A_{f},A^{\dagger}_{g}] = (f,g) = [B_{f},B^{\dagger}_{g}], 
   \quad [A_{f},B_{g}] = 0, \quad [A^{\dagger}_{f},B^{\dagger}_{g}] = 0 ~, 
  \lab{p518}\ee 
  with $(f,g)$ denoting the scalar product between $f$ and $g$. Now  
  $<A_{f}|A_{f}> = 1$ and the $A_{f}$'s are well defined operators 
  in the Fock space where observables have to be 
  realized.  
  The $A_{f}$ number operator is then 
  \be 
  {n}_{A_{f}}(t) = < 0(\theta,t) | A_{f}^{\dagger}(\theta) A_{f}(\theta)  
  | 0(\theta,t) > = 
   {1 \over{(2\pi)^{3}}}\int {d^3{ k}} 
  \sinh^{2} \bigl( \Gamma_{ k} t \bigr) |f({ k})|^{2} 
  ~ , 
  \lab{p2746}\ee 
  and similarly for the modes of type $B_{f}(\theta)$ (cf. with  
  eq. (\ref{p274})). We can set ${n}_{A_{f}}(t)  
  \equiv \sinh^{2} \bigl( \Gamma t \bigr)$ and Eq. (\ref{p2746}) then  
  specifies the relation between the  
  ${\Gamma}_{ k}$'s and ${\Gamma} \equiv H/2$. 
   
  The structure of $|0({\theta},t)>$ naturally leads us to recognize  
  its thermal properties.  
  The vacuum state $|0({\theta},t)>$ can be  
  written as 
  \be 
  |0({\theta},t)>\, = \exp{\left ( - {1\over{2}} {\cal S}_{A({\theta})}  
  \right )} |\,{\cal I}({\theta})>\, = \exp{\left ( - {1\over{2}} {\cal  
  S}_{B({\theta})} \right )} |\,{\cal I}({\theta})> \quad , 
  \lab{p81}\ee 
  where 
  $ 
  |\,{\cal I}({\theta})>\, \equiv \exp {( \sum_{ k} 
  A_{ k}^{\dagger}({\theta}) 
  B_{ k}^{\dagger}({\theta}) )} |0({\theta})>$ is 
  the invariant (not normalizable) vector ~\cite{TFD} 
  and 
  \be 
  {\cal S}_{A({\theta})} \equiv - \sum_{ k} \Bigl \{ 
  A_{ k}^{\dagger}({\theta}) A_{ k}({\theta})  
  \ln \sinh^{2} \bigl ( \Gamma_{ k} t \bigr ) - A_{ k}({\theta})  
  A_{ k}^{\dagger}({\theta}) \ln \cosh^{2} \bigl ( \Gamma_{ k} t 
  \bigr ) \Bigr \}~~. 
  \lab{p83}\ee 
   
  ${\cal S}_{B({\theta})}$ has the same expression  
  with $B_{ k} 
  ({\theta})$ and $B_{ k}^{\dagger}({\theta})$ 
  replacing $A_{ k}(\theta)$  and 
  $A_{ k}^{\dagger}({\theta})$,  
  respectively.  In the following we  
  shall simply write ${\cal S}({\theta})$ for either ${\cal  
  S}_{A({\theta})}$ 
  or ${\cal S}_{B({\theta})}$. 
  ${\cal S}({\theta})$  is recognized to be the entropy \cite{QD, TFD}.
   
  Since the $B$-particles are the holes for the $A$-particles, ~~ 
  ${\cal S}_{A(\theta)} - 
  {\cal S}_{B(\theta)}$~~ is in fact the (conserved) entropy for the closed 
  system: 
  $ 
  [\, {\cal S}_{A(\theta)} - {\cal S}_{B(\theta)} , {\cal H}  
  ] = 0$ .

  Eqs. (\ref{p81}) and (\ref{p83}) show that the operator  
  dependence of ${1\over{2}} 
  {\cal S}_{A({\theta})}$ 
  (or respectively,  ${1\over{2}} {\cal S}_{B({\theta})}$) is  
  uniquely on the  
  $A$ ($B$) variables: thus in 
  eq. (\ref{p81}) time evolution is expressed solely in terms of the  
  (sub)system $A$ ($B$) 
  with the elimination of the $B$ ($A$) variables. This reminds us of the 
  procedure by which one obtains the reduced density matrix by integrating 
  out bath variables. 
   
  For the time variation of $|0(\theta ,t)>$ at finite volume $V$, we obtain 
  \be 
  {{\partial}\over{\partial t}} |0(\theta ,t)> =  -  {1\over{2}} \left (  
  {{\partial {\cal S}({\theta})}\over{\partial t}} \right ) 
  |0(\theta ,t)>  \quad . 
  \lab{p91}\ee 
   
  Equation (\ref{p91}) shows that $ {1\over{2}} \left (  
  {{\partial {\cal S}({\theta})}\over{\partial t}}  \right ) $ is the 
  generator of time-translations, 
  namely time evolution is controlled by the entropy variations.   
  This correctly reflects the 
  irreversibility of time evolution characteristic of expanding metrics.
  Expanding geometry implies in fact the choice of a privileged 
  direction in time evolution ({\it time arrow}) with a consequent breaking of 
  time-reversal invariance. 
   
  Let us now  consider the $A$-modes alone and introduce the  
  functional (free energy) 
  \be 
  F_{A} \equiv <0({\theta},t)| \Bigl ( {{\cal H}^{\prime} }_{0,A(\theta) } - 
   {1\over{\beta}} {\cal S}_{A({\theta})} \Bigr ) |0({\theta},t)>.  
  \lab{p92}\ee 
  Here ${{\cal H}^{\prime}}_{0,A(\theta)} \equiv \sum_{ k} E_{ k}  
  A_{ k}^{\dagger}(\theta) A_{ k}(\theta)$ and 
  $E_{ k} \equiv 
  \hbar \Omega_{ k}(t_{0}=0) - \mu$, with $\mu$ the chemical 
  potential. The stability condition  
  ${{\partial {F}_{A(\theta)}}\over{\partial \sigma_{ k}}} = 0, \;   
  {\sigma}_{ k} \equiv \Gamma_{ k} t$ $\forall  k,$  
  assuming $\beta$  a slowly varying  
  functions of t, gives 
  $\beta E_{ k} = - \ln \tanh^{2} \bigl ( \sigma_{ k} \bigr ) $, i.e. 
  \be 
  {n}_{A_{ k}}(t) = \sinh^{2} \bigl ( \Gamma_{ k} t \bigr ) =  
  {1\over{{\rm e}^{\beta (t) E_{ k}} - 1}} \quad , \lab{p101} 
  \ee 
  which is the Bose distribution for $A_{ k}$ at time $t$ provided 
  we assume $\beta (t)$ to represent the inverse temperature  
  $\beta(t) = {1\over{k_{B} T(t)}}$ at time $t$ ($k_{B}$ denotes the Boltzmann 
  constant).  This allows us to recognize $\{ |0(\theta ,t)> \}$ as a  
  representation of 
  the canonical commutation relations at finite temperature, equivalent 
  with the Thermo Field Dynamics representation $\{ |0(\beta )> \}$ of  
  Takahashi and Umezawa ~\cite{TFD, Um1, Um} . See also 
  \cite{Double, Jagna}. 
   
  Let us now finally comment on the reality condition for  
  $\Omega_{k}(t)$. 
  Our  
  first remark is that 
  such a condition actually  
  excludes long wave modes, and thus acts as an 
  intrinsic infrared cut-off; in fact, it is easy to show that  
  ${\Omega_{k}}^{2}(t) \ge 0$ for any $t$ implies  $k
  \ge k_{0} e^{{{H \over {3}}t}}$, with $ k_{0} \equiv {Ha_{0} 
  \over  {2c}}$
    for  any  
  $t$.  We recover in this way 
  the known feature of inflationary models by which only in  
  the ``tight coupling'' phase 
  ($\lambda \,<\, R_{H}$) there is an oscillatory evolution\cite{Al}. 
  
  As a matter of fact, besides the reality condition,  
  we also have the condition (\ref{c0})  which implies that  
  ${\Omega_{k}}^{2}(t) \le {\omega_{0}}^{2}$ for $\theta_{k} \ge 0$ and that 
  ${\Omega_{k}}^{2}(t) \ge {\omega_{0}}^{2}$ for $\theta_{k} \le 0$.  
  This, together with the reality condition, leads to the bounds for  
  $k$:  

  \be 
  k_{0} e^{{H \over {3}}t} \le k \le {\tilde k}_{0}   
  e^{{H \over {3}}t} 
    ~~~ at~~~ any~~~t~~~for~~~ 
  {\theta}_{k} \ge 0~, 
  \lab{c2}\ee  
  \be 
   k \ge {\tilde k}_{0}   
  e^{{H \over {3}}t} 
    ~~~ at~~~ any~~~t ~~~for~~~ 
  {\theta}_{k} \le 0~, 
  \lab{c4}\ee  
  with ${\tilde k}_{0} \equiv  
   \frac{a_{0}}{c}\,\sqrt{\omega^{2}_{0} +  
  \left(\frac{H}{2}\right)^{2}}$. 

  On the other hand, {\it for each given mode $k$}, the frequency  
  $\Omega_{k}$ is different from zero only in the span of time  
  $0 \le t \le {\rm T}_{k} \equiv  
  {3 \over {H}}\ln {k \over {k_{0}}}$  
  (limiting ourselves to positive time evolution).  
  For instance, for each $k$  we have:   
  \be {\Omega_{k}(\Lambda_{k}(t))}\,=\,  \Omega_{k}(0) e^{-\Lambda_{k}(t)}~, 
  ~~ ~\Lambda_{k}(t) \ge 0 ~~for~~any~~t,
  \lab{c5}\ee 
  with 
  \be 
   e^{-2\Lambda_{k}(t)} \,\equiv\,  
  \frac{e^{-t \frac{H}{3}}\,{\rm sinh}\frac{H}{3}({\rm T}_{k}-t)}{{\rm sinh} 
  \frac{H}{3}{\rm T}_{k}} \qquad , 
  \lab{c6}\ee 
  and ${\Omega_{k}(\Lambda_{k}(0))}\,=\,  \Omega_{k}(0) $  and 
  ${\Omega_{k}(\Lambda_{k}({\rm T}_{k}))}\,=\,0$.  
  Modes with larger $k$ have "longer" life with reference to time $t$. 
  Asymptotically in $k$, the limit value $\Lambda_{asimpt.}\, 
  =\, \frac{H}{3}\, t$ is reached (see figs.1 and 2).  
  In conclusion, only the modes
  satisfying conditions (\ref{c2}) and (\ref{c4}) are present at time $t$,
  being the other ones decayed (fig. 2).
  
  \begin{figure}[h]
  \epsfig{file=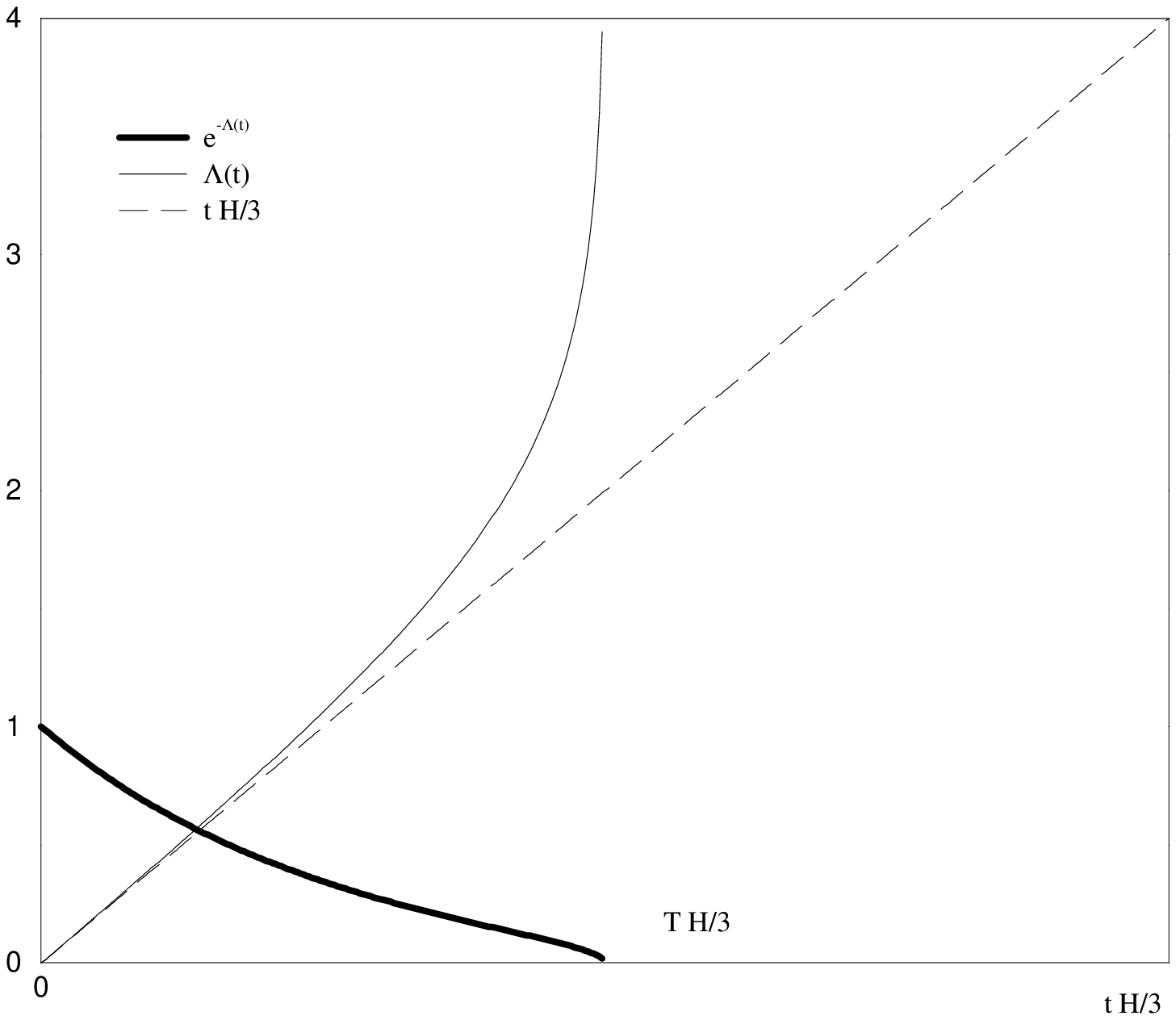,width=1.0\linewidth,height=0.8\linewidth}
  \lab{fig1}
  \end{figure}
  \centerline{fig. 1}
  \bigskip
  \begin{figure}[t]
  \epsfig{file=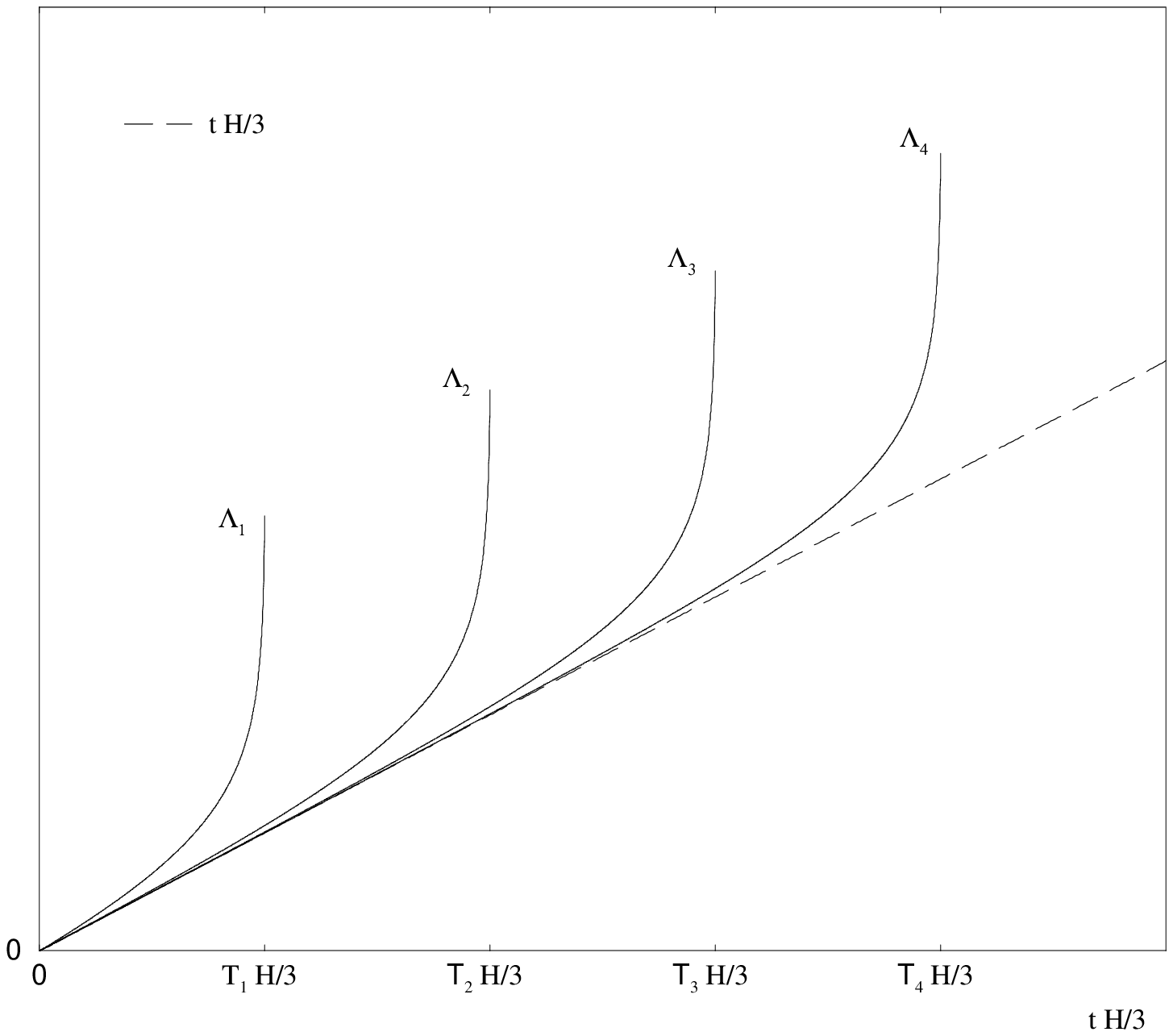,width=1.0\linewidth,height=0.8\linewidth}
  \end{figure}
  \smallskip
  
  The condition (\ref{c0}) 
  allows to write also 
  \be {\Omega_{k}(t)}\,=\,  \omega_{0} e^{-\theta_{k}(t)}~.
  \lab{c7}\ee 
  Thus we see from (\ref{c5}) and (\ref{c7}) that $\Lambda_{k}(t) =  
  \theta_{k}(t)  - \theta_{k}(0)$, i.e. defining 
   $\Lambda_{k}(t)\,=\, \gamma_{k} \tau_{k}(t)$
  ($\gamma_{k}\tau_{k}(0)=0$, 
  $\gamma_{k}\tau({\rm T}_{k})=\infty$), we have
   \be {\Omega_{k}(t)}\,=\,  \omega_{0} e^{-\theta_{k}(0)}
   e^{- \gamma_{k} \tau_{k}(t)}  ~,
  \lab{c71}\ee 
  which shows that 
  $\tau_{k}(t)$ can be interpreted  as the {\it proper} time of 
  the $k$-mode.

  It is finally interesting to remark that the number $n_{k}$ of  
  $k$-modes condensed in the state ~$|0(\theta)>$,~ given by  
  $\sinh^{2}{\theta_{k}}$, can be expressed as ~ 
  ${n_{k}\,\equiv\,n_{+,k}+n_{-,k}}$~
  in terms of

\centerline{fig. 2}
\centerline{}
\noindent  the $k$-modes  
  $n_{+,k}$ and $n_{-,k}$  
  going forward and backwards in the proper time $\tau_{k}$, 
  respectively:
  \be 
  n_{\pm,k}(t)=\frac{n_k(t)}{ 
  \ e^{\mp 2 \theta_{k}(0)}  
  e^{\mp 2 \gamma_{k} \tau_{k}(t)} +1} 
  \ee 
   
  Clearly, from (\ref{c0}) one recognizes that  
  $n_{+,k} - n_{-,k}$ can be considered as an "order parameter" 
  since  
  $\tanh{\theta_{k}(t)} =  \frac{n_{+,k}(t) -  
  n_{-,k}(t)}{n_{k}(t)}$~, and $n_{+,k} = n_{-,k}$ for any $t$ 
  for each $k$-mode with constant frequency ($\theta_{k} = 0$  
  for any $t$). 
    
  This work has been partially supported by INFN, by MURST  
  and by a Network of the European Science Foundation  
  on Topological Defects.

  \end{document}